# Constraining the pitch angle of the galactic spiral arms in the Milky Way

Jacques P. Vallée

National Research Council of Canada, National Science Infrastructure, Herzberg Astronomy & Astrophysics, 5071 West Saanich Road, Victoria, B.C., Canada V9E 2E7

**Abstract.**

We carry out analyses of some parameters of the galactic spiral arms, in the currently available samples.

We present a catalog of the observed pitch angle for each spiral arm in the Milky Way disk. For each long spiral arm in the Milky Way, we investigate for each individual arm its pitch angle, as measured through different methods (parallaxes, twin-tangent arm, kinematical, etc), and assess their answers.

Second, we catalog recent advances in the cartography of the Galaxy (global mean arm pitch, arm number, arm shape, interarm distance at the Sun). We statistically compare the results over a long time frame, from 1980 to 2017. Histograms of about 90 individual results published in recent years (since mid-2015) are compared to 66 earlier results (from 1980 to 2005), showing the ratio of primary to secondary peaks to have increased by almost a factor of 3. Similarly, many earlier discrepancies (expressed in r.m.s.) have been reduced by almost a factor 3.

## 1. Introduction

In this contribution, we present exploratory analyses of galactic spiral arms properties, notably the pitch angle, with the aim of constraining their individual and global values, as best could be done with currently available samples.

Our knowledge of the main parameters of the spiral arms (their number, their shape, their pitch angle, and the interarm separation through the Sun between the Sagittarius and the Perseus arms) has evolved with time, but some discrepancies have lingered on.

An early picture of the location of each spiral arm is that in Courtès et al (1969 – their fig. 6 and Table 2), with 4 arms, a pitch angle of -20$^{o}$, an interarm separation of about 4kpc, and a approximate log shape (disregarding the local Orion armlet), using 10 kpc for the distance from the Sun to the Galactic Center. A very recent model picture can be seen in Fig. 2 of Vallée (2016a), with 4 arms, a pitch angle of $-13^o$, an interarm separation of about 3kpc, and a logarithmic arm shape, using 8 kpc for the Sun to Galactic Center distance.

The “twin-tangent” method employs an ideal model of a spiral arm with parallel layers, each layer would contain a different tracer (dust, maser, CO in an arm). The layer closest to the direction of the Galactic Center (GC) is the hot dust tracer. A look from Earth at a tangent to an arm, looking in one layer/tracer, would give an angle from the GC (the galactic longitude of that tracer in that arm). Employing a given tracer (cold dust, say) in a long arm, on each side of the GC (in Galactic Quadrant I and IV), the two different tangent angles measured could be fitted to deduce the arm's pitch angle. Doing the same using any another tracer (CO, say), would give a similar result, hence showing very little differences (a modest spread around a mean pitch angle). The twin-tangent method uses the two tangents to the same arm, as observed on both side of the sun-galactic center line in Galactic Quadrants IV and I (equation 1 in Drimmel 2000, or equation 10 in Vallée 2015).

The “parallax” method looks at a slice of an arm, namely the arm's inner side closest to the GC (where all masers are located). By measuring a maser's distance from Earth, and that of neighboring masers, these masers can be plotted on the galactic plane (longitude and distance from earth). Next, a straight line can be fitted through the data (masers) representing the arm, and the line's pitch angle (arm pitch) is the angle away from a circle around the Galactic Center. Projecting an arm from a few maser locations may lead to predicting different spiral arcs, and different predicted tangents to a spiral arm. Masers are also found in short spurs or armlets, growing out of a long arm; each maser paper focuses on a specific piece of sky. Doing the same pitch angle deduction at other data located far away along the same arm (different galactic longitudes), should give a similar result, showing very little differences (a little spread around a mean pitch angle).

The “kinematic” method assumes a velocity model to obtain distances from the Sun, while the “luminosity-distance” method assumes a dust absorption model with distances. Both are employed to position the observed objects on the Galactic plane, after which a pitch angle fit is extracted for the spiral arm involved.

The “positional” method extracts observational arm values (arm number, arm shape, arm pitch angle, arm separation near the Sun) from fits to different individual objects (stars, masers, HII regions, etc) positioned on the Galactic plane. In this paper, we look for signs of convergence over time (a shallower width in the distribution, a higher primary peak in the distribution).

In Sections 2 and 3, we aim to assess each spiral arm’s pitch angle. Due to the inherent differences in the nature of these methods, there is a concern as to which would give the more precise determination of pitch angle. Section 4 catalogues results published since 2015, using the positional method (arm number, arm shape, arm pitch, and interarm separation near the Sun). In Section 5, the results are assessed over time, to evaluate convergences. In Section 6 we assess the results since mid-2015, comparing to those done in a 2005 study. In Section 7, we employ a proper galactic spiral arm pitch value to present a cartographic and kinematic model of the Milky Way. We conclude in Section 8.

## 2. Individual arm pitch angle

Each arm can now be identified by tracers, placed in a specific order and at specific galactic longitudes. A recent study of the galactic longitude of each arm tangent (as seen from the Sun) showed longitude offsets between dust, stars and various chemical tracers such as CO (Paper VI; Vallée 2014c). Going across galactic longitude $0^o$, the galactic longitudes of the tracers (CO, then dust) in Galactic Quadrant IV reversed as one went across the Galactic Meridian to Galactic Quadrant I (dust, then CO) - see fig. 1 in Vallée (2016b).

Some pitch angle values have been found for portions of individual arms, in the Milky Way. Many pitch angle values for portions of arms were deduced from maser surveys. There is the BeSSel maser survey (Brunthaler et al 2011, Zhang et al 2013; Choi et al 2014; Reid et al 2014; Sato et al 2014; Wu et al 2014; Zhang et al 2014), and there is the VERA maser survey (Sato et al 2010; Nagayama et al 2011; Sakai et al 2012; Honma et al 2012; Chibueze et al 2014; Nakanishi et al 2015; Sakai et al 2015). Pitch angle values found for arm portions may differ, but still around the mean global pitch angle for a long arm in the Milky Way.

### 2.1 Data

**Table 1** shows the pitch angle for each individual spiral arm, as taken from the literature.

The twin-tangents method was used for the Carina arm attached to the Sagittarius arm, the Crux-Centaurus arm attached to the Scutum arm, and the 'Start of Norma' arm attached to the Norma arm. The mean arm pitch angle across all spiral arms from the twin-tangent method is -13.6 ±0.4 degrees.

Using the 'parallax' method, Krishnan et al (2017) proposed to place the high-mass star forming region G305.2 in the Carina arm, rather than the Crux-Centaurus arm (their Section 6.2), using parallaxes of methanol masers. Parallax-derived arm pitch angle values vary from $2^o$ to $20^o$. The mean arm pitch angle across all spiral arms from the parallax method is -12.3 ±2.2 degrees.

### 2.2 Summary

As can be seen, the twin-tangent method has a mean pitch angle with a small r.m.s. of only 0.4$^o$, while the parallax method has a mean pitch angle with a r.m.s. of 2.2$^o$.

An average of these two methods gives a mean pitch angle near -13$^o$.

## 3. Histograms of the individual pitch angle, for three inner arms

The data in Table 1 can be employed to make histograms, for each arm, and for an individual method.

### 3.1 Sagittarius-Carina arm

**Figure 1** shows, for the Sagittarius-Carina spiral arm, histograms of measured pitch angles.

Fig. 1a shows all data from four methods (parallax, twin-tangents, kinematical and luminosity).

Fig. 1b shows only the data from the twin-tangent arm method. The distribution of pitch angle goes from -12$^o$ to -15$^o$.

Fig. 1c shows only the data from the parallax method. The distribution of pitch angle goes from -6$^o$ to -19$^o$.

### 3.2 Scutum-Crux-Centaurus arm

**Figure 2** shows, for the Scutum-Crux-Carina arm, histograms of measured pitch angles.

Fig. 2a shows all data from four methods (parallax, twin-tangents, kinematical and luminosity).

Fig. 2b shows only the data from the twin-tangent arm method. The distribution of pitch angle goes from -11$^o$ to -15$^o$.

Fig. 2c shows only the data from the parallax method. The distribution of pitch angle goes from -6$^o$ to -20$^o$.

### 3.3 The Norma arm

**Figure 3**a shows all data from two methods (twin-tangent, kinematical).

Figure 3b shows only the data from the twin-tangent arm method. The distribution of pitch angle goes from -11$^o$ to -15$^o$.

There are no data from the parallax method for the Norma arm. **Figure 3c** shows the data from the parallax method for the extension of the Norma arm in Galactic Quadrant II (the so-called Cygnus arm). The distribution of pitch angle goes from -2$^o$ to -15$^o$. A sketch of this long arm is shown in red in Vallée (2016a – fig.2).

### 3.4 Summary

As are seen in these histograms, for each arm, the width of the distribution in the histograms of pitch angle values is only $4^o$ for the twin-tangent method (Fig. 1b, 2b, 3b), while it is near $13^o$ for the parallax method (Fig. 1c, 2c, 3c).

The pitch angle observed in many nearby disk spiral galaxies appears approximately flat over 10 kpc, save for localized deviations with an amplitude near $20^o$ - a close look at the galaxy M51, using both optical data and radio data, confirms this view (Vallée 2016a, his fig. 1). A $20^o$ deviation from the mean pitch angle is about 11% of the full $180^o$ range of the mean. Similarly for the Milky Way, one could expect that each arm has an approximate mean pitch angle value with radial distance from the galactic center (except for small localized deviations), and that this mean value is similar to the mean pitch angle value of any other spiral arm (to first order).

## 4. Positional method

In a series of papers, we have catalogued the published observational results since 1980 for the Milky Way's arms (number of arms, arm shape, pitch angle, interarm separation through the Sun's location). Results were put in blocks, each block with a minimum of 15 and a maximum of 20 results. Papers in this series were: Vallée (1995 – Paper I), Vallée (2002 – Paper II), Vallée (2005 – Paper III), Vallée (2008 – Paper IV), Vallée (2013 – Paper V), Vallée (2014a – Paper VI), Vallée (2014b – Paper VII), Vallée (2015 – Paper VIII), Vallée (2016a – paper IX).

**Table 2** here show the most recent published papers on this topic (Paper X), using the same maximum and minimum number of results per bloc. Fitted data (pitch angle, interarm separation at the Sun's position, arm shape and arm number) are taken from their original published figures, employing a global model (their own, or one they adapted themselves from the literature).

The weight of each study in Table 2 is indicated as a full fit (strong) or a light fit (partial), in column 6. We omitted studies where authors provide a figure but no fit (illustrative), showing the place of an object on a previously published plot. The statistical reliability of the results and the methods employed, as well as the technical accuracy of the results, have been shown earlier to give overall means and standard deviations that are relatively unchanged when using different weights or equal weight, for these types of galactic arm studies (see Table 2 in Paper I; Tables 1 and 2 in Paper III; Table 1 in Paper IV).

The last rows showed the most recent statistics; the results do not diverge significantly from the previously published ones.

In the Appendix below, we briefly discuss some comments on these measurements.

**5. Statistical trends with time, since 1980**

Here we wish to assess the evolution of our knowledge with time of some spiral arm parameters in the Milky Way disk, over the period from 1980 to 2017.

The median value of the observed pitch angle data since 2015 is near $-13^{o}$ with a r.m.s. near $0.5^{o}$, for the positional method (Table 2). Earlier data indicated a mean near $-12^{o}$ with an r.m.s. near $1^{o}$ (Vallée 2005).

The value of the observed interarm separation since 2014 is near 3.1 kpc with a r.m.s. near 0.1 kpc. Earlier data indicated a mean near 3.4 kpc with an r.m.s. near 0.3 kpc (Vallée 2005).

The percentage of the studies having fitted a logarithmic arm shape since 2014 is near 94%. Earlier data indicated a percentage near 88% (Vallée 2005).

The percentage of the studies having four spiral arm shape since 2014 is near 87%. Earlier data indicated a percentage near 80% (Vallée 2005).

**6. Statistical convergence of recent results**

Here we assess the histograms of individual spiral arm results, for the published studies since 2014 (covering Paper VIII, IX and X). We wish to assess whether we have a single peak or not, and its importance.

**Figure 4a** shows a histogram from a compilation of the observed pitch angle. Each data represents one individual published study; there are 94 such data since 2014. The central peak here is about eight times higher than the adjacent secondary peak, whereas earlier it was only about three times higher (Fig. 3a in Vallée 2005) in 66 studies published before mid-2005. Thus the peak has improved by a factor near 3.

A histogram (**Fig. 4b**) of the observed interarm separation between the Sagittarius and the Perseus arms (crossing the Sun's position), covering 90 studies since 2014, shows a central peak about fifteen times higher than the adjacent secondary peak, whereas earlier it was only about six times higher (Fig. 3d in Vallée 2005) in 66 studies published before mid-2005. Thus the peak has improved by a factor near 3.

A histogram (**Fig. 4c**) of the spiral arm shape observed, covering 95 studies since 2014, shows a central peak about forty times higher than the next secondary peak, whereas earlier it was only about sixteen times higher (Fig. 3c in Vallée 2005) in 66 studies published before mid-2005. The peak has improved by a factor near 3.

A histogram (**Fig. 4d**) of the number of spiral arms observed, covering 91 studies since 2014, shows a central peak about sixteen times higher than the next secondary peak, whereas earlier it was only about five times higher (Fig. 3b in Vallée 2005) in 66 studies published before mid-2005. The peak has improved by a factor near 3.

None of the histograms currently shows a single narrow peak, but all of them now show a strong central peak towering above other lesser secondary peaks. The ratio of the primary peak to the secondary peak has increased by a factor near three, over the last twelve years, hence a good sign of a greater convergence of these arm parameters.

**7. Modeling**

A review of around 50 determinations of $R_{sun}$ , published between 1992 and 2011, found a weighted mean value of 8.0 ± 0.4 kpc, covering a 20-year time period (Fig. 1 in Malkin 2013). A review of about 70 determinations of $R_{sun}$ , published from 1990 up to mid-2012, was given in Gillessen et al (2013), and their Fig. 2 showed a median near 8.1 ±0.3 kpc. The review of $R_{sun}$ by De Grijs & Bono (2016), covered 273 entries since 1918, and yielded $R_{sun}$ = 8.3 ±0.4 kpc. A review by Bland-Hawthorn & Gerhard (2016) of 26 determinations, published with data from 2005 to 2015, yielded an unweighted arithmetic mean of 8.0 ±0.4 kpc (their Table 3). A review of 27 recent determinations (from mid-2012 to early 2017) by Vallée (2017a – his table 1) gave 8.0 ± 0.2 kpc. Here we take this value of 8.0 kpc in our velocimetric model.

Recent published measurements of the orbital circular velocity of the Local Standard of Rest (LSR) around the Galactic Center appear to show a grouping, one around 210-230 km/s, and another one around 230-250 km/s. A review of 24 recent determinations (from mid-2012 to early 2017) by Vallée (2017a – his table 1) gave a weigted value of 229 ± 3 km/s and a median value of 232 km/s. Hence we take a value of 230 km/s to employ in our velocimetric model.

In recent papers, we obtained accurate values for the galactic longitudes of the observed arm tangents, for each different arm tracer (Vallée 2014a; Vallée 2014c; Vallée 2016b). The origin of the Sagittarius arm in Galactic quadrant IV, and the origin of the Norma arm in Galactic quadrant I (various arm tracers in Table 4 in Vallée 2016a), and their separation of 2.2 kpc from the GC (fig. 4 in Vallée 2016a) were recently determined. Hence we take a separation of 2.2 kpc to employ in our velocimetric model.

As done in earlier models, and given that most observations found 4 arms, we employ here a 4-arm model, as well as the logarithmic model equations listed earlier in Vallée (2008). However, more accurate input parameters are employed here.

A revised velocimetric model is computed here, employing the parameters $R_{sun}$ = 8.0 kpc; four arms equally spaced in azimuth, with arm pitch= $-13.1^{o}$, showing two circular orbital velocity values around the mean of 230 km/s. As often done, we employ a mean arm in our model to represent each arm.

**Figure 5** shows a rough cartographic and kinematical model, with a proper global pitch angle, showing all the arms named in Table 1 (including the Cygnus + I arm). **Figure 5a** shows the best cartographic map to date, for the Outer Milky Way ($2^{nd}$ and $3^{rd}$ galactic quadrants). **Figure 5b** shows the best velocimetric map to date, for the Outer Milky Way ($2^{nd}$ and $3^{rd}$ galactic quadrants).

**Table 3** shows a selected sample of often-used velocimetric models as published earlier, giving the parameters of the spiral arms employed.

While these earlier velocimetric models resulted in some good overlaps (see Zucker et al 2015 for the Scutum arm), a more precise model would use the more precisely determined spiral arm values, including the mean pitch angle for the spiral arms.

Similar cartographic figures were produced for the other galactic quadrants IV and I - see Vallée 2017c (fig.3 - following the Norma arm); Vallée 2016a (fig.2 - near the GC).

Similar velocimetric figures were produced for the other galactic quadrants IV and I - see Vallée 2017c (Fig.4 - following the inner Norma arm); Vallée 2017b (Figs 1 and 2 - within $30^{o}$ of the GC).

Reid et al (2016) proposed a Bayesian computer routine to draw spiral arms; they predicted a Carina arm tangent at galactic longitude $l=302^{o}$ (their Figure 5b; see also their model as reproduced in Fig.1 in Ragan et al 2016), and this longitude is quite different than observations from longitude $281^{o}$ (CO) to $285^{o}$ (dust) – see Vallée (2016b – table 3). This discrepancy (their prediction versus observations) is significant ($17^{o}$ at 5 kpc from the Sun is 1 450 pc).

The start of the Norma arm is crucial to separate some kinematic models. Fig. 4 in Green et al (2017) showed for the Norma arm in Galactic Quadrant I the arm tangent longitudes ($16^{o}$ to $20^{o}$ for the chemical tracers listed in Vallée 2016b), and their kinematic model for the Norma arm (from $0^{o}$ to $10^{o}$). Their unexplained gap between the two (from $10^{o}$ to $18^{o}$) is due to their model Norma arm starting at 3.1 kpc from the Galactic Center, instead of starting at 2.2 kpc from the Galactic Center (see Fig. 3 and Fig. 4 in Vallée 2017c; also Fig. 4 in Vallée 2016a and Fig. 2 in Vallée 2017b).

## 8. Conclusion

By linking each arm segment in Galactic Quadrant I with its corresponding arm segment in Galactic Quadrant IV, simple trigonometry using a logarithmic shape reveals the *mean pitch angle* of each arm. We compared the pitch angle of individual arms (Table 1), assessing the pros and cons of the parallax method versus the twin-tangent arm method (Figures 1, 2, 3).

We find the twin-tangent method to give a more credible ‘global arm’ pitch angle value, based on data from a larger amount of galactic longitudes. We then use statistics from the twin-tangent arm method to provide a ‘global galactic’ pitch angle value, averaging over individual spiral arms, giving $-13.6^{o} \pm 0.4^{o}$.

For each arm, the width of the distribution in the histograms of pitch angle values is near $4^{o}$ for the twin-tangent method.

Then we statistically analyse catalogues of recent global observed properties of the spiral arms (Table 2), as obtained *simultaneously*. They are given in the same manner and format as in previous papers I to IX in this series. We employed these results to do some analyses over nearly four decades, to look for trends and possible convergences, and to compare with an earlier similar study done in 2005 (Paper III).

An histogram of published data since 2014 show a clearer picture for the mean pitch angle (Fig. 4a). There is a good convergence over time: a shallower width in the distribution by a factor near 3, a higher primary peak in the distribution by a factor near 3 since the study done in 2005. Similar histograms show improvements for the interarm separation (Fig. 4b), arm shape (Fig.4c), and number of arms (Fig. 4d).

These global properties are useful and needed when modelling spiral arms, for both cartographic (Fig. 5a) and velocimetric (Fig. 5b) studies.

**Acknowledgements.**

The figure production made use of the PGPLOT software at NRC Canada in Victoria. I thank an anonymous referee for useful, careful, and historical suggestions.

## Appendix

Here we briefly discuss some earlier measurements.

Melnik et al (2016a) assumed two rings, with the Sun set at a galactic radius near 7.5 kpc. Their model tangent for the ‘ascending’ segment of R1 (Carina) is at a galactic longitude of $+302^{o}$ (their fig. 2b and fig. 4a), but the observed Carina arm tangent is near $l= 283^{o}$ (Table 1 in Vallée 2014c). Their ring R1 has an ‘ascending’ segment of pitch angle $+6^{o}$ (their Fig. 2b and Fig. 4a) for the Carina arm, and a ‘descending’ segment of pitch angle $-6^{o}$ (their Fig. 2b and Fig.4a) for the Sagittarius arm.

Koda et al (2016) employed a long thin stellar and molecular bar, of radius 4.4 kpc at position angle $44^{o}$ to the Sun-Galactic Center line. They started their Sagittarius arm near longitude $l = -9^{o} = 351^{o}$ at 3 kpc from the Galactic Center, but the Sagittarius arm tangent is observed at $l = -17^{0} = 343^{o}$ near 2.2 kpc (Vallée 2016a). They started the Norma arm near $l= +8^{o}$ at 3 kpc from the Galactic Center (their fig.11), while the observed Norma arm tangent is observed at/near $l= +20^{o}$ (Vallée 2016a).

**Table 1 – Observed individual pitch angle (p, in degrees, negative inward), for each spiral arm in the Milky Way galaxy**

| Start of Perseus arm | Norma arm | Scutum-Crux-Centaurus arm | Carina-Sagitt. arm | Perseus arm | Cygnus arm | Cygnus + 1 arm | Method[(a)] | Data used[(b)] | Reference [(c)] |
|---|---|---|---|---|---|---|---|---|---|
| l= 337$^o$ - | l=329$^o$ l= 20$^o$ | l=309$^o$ l= 33$^o$ | l=281$^o$ l= 51$^o$ | outer Galaxy | outer Galaxy | outer Galaxy | | | |
| p | p | p | p | p | p | p | | | |
| - | - | -12.4 | -12.4 | -12.4 | -12.4 | - | kin | HI gas | Tab.1 and Fig.6 in Koo et al (2017) |
| - | - | - | -19.0 | - | - | - | par | meth. masers | Sect.6.2 in Krishnan et al (2017) |
| - | -11.1 | - | - | - | - | - | tan | CO at 8' | Tab.1 in Vallée (2017c) |
| - | -15.8 | - | - | - | - | - | tan | synchr. 408MHz | Tab. 1 in Vallée (2017c) |
| - | -14.3 | - | - | - | - | - | tan | masers | Tab.1 in Vallée (2017c) |
| - | - | - | - | - | -13.1 | - | kin | CO gas | Fig.2 in Du et al (2016) |
| - | -15 | -11 | -11 | -15 | -15 | - | kin | HI and CO | Tab. 1 in Nakanishi & Sofue (2016) |
| - | - | - | -14.5 | - | - | - | tan | CO at 8' | Tab.1 in Vallée (2015) |
| - | - | - | -15.1 | - | - | - | tan | thermal electron | Tab.1 in Vallée (2015) |
| - | - | - | -13.5 | - | - | - | tan | old HII complex | Tab.1 in Vallée (2015) |
| - | - | - | -14.1 | - | - | - | tan | dust | Tab.1 in Vallée (2015) |
| - | - | - | -12.6 | - | - | - | tan | FIR [CII] & [NII] | Tab.1 in Vallée (2015) |
| - | - | -11.9 | - | - | - | - | tan | CO at 8' | Tab.2 in Vallée (2015) |
| - | - | -12.7 | - | - | - | - | tan | thermal electron | Tab.2 in Vallée (2015) |
| - | - | -12.7 | - | - | - | - | tan | old HII complex | Tab.2 in Vallée (2015) |
| - | - | -14.3 | - | - | - | - | tan | FIR [CII] & [NII] | Tab.2 in Vallée (2015) |
| - | - | -12.2 | - | - | - | - | tan | Synchr. 408 MHz | Tab.2 in Vallée (2015) |
| - | - | -14.5 | - | - | - | - | tan | HI | Tab.2 in Vallée (2015) |
| - | - | --13.8 | - | - | - | - | tan | dust | Tab.2 in Vallée (2015) |
| - | - | -19.2 | - | - | - | - | par | meth. masers | Sect.5.2 in Krishnan et al (2015) |
| - | - | - | - | -11.1 | - | - | par | $H_2O$ masers | Fig.4 in Sakai et al (2015) - VERA |
| - | - | -10.0 | -10.5 | -7.9 | -10.3 | - | lum | Cepheids | Table 1 in Dambis et al (2015) |
| - | - | - | - | - | - | -9.3 | kin | CO | Fig.3 in Sun et al (2015) |
| - | - | - | - | - | -14.9 | - | par | $H_2O$, meth. masers | Fig. 6 in Hachisuka et al (2015) |
| - | -9.9 | -10.5 | -10.0 | -8.1 | -2.7 | - | kin | HII and GMC | Tab. 1 in Hou & Han (2014) |
| - | - | -19.8 | - | - | - | - | pal | $H_2O$, meth.masers | Fig. 4 in Sato et al(2014) - BeSS |
| -5.6 | -6.6 | -13.4 | - | - | - | - | kin | CO clouds | Tab. 3 in Garcia et al (2014) |
| - | - | -19.8 | -6.9 | -9.4 | -13.8 | - | par | $H_2O$, meth. masers | Tab. 2 in Reid et al(2014)-BeSS |
| - | - | -11.2 | -9.3 | -14.8 | -11.5 | - | par | $H_2O$, meth.masers | Tab.2 in Bobylev & Bajkova (2014) |
| - | - | - | -7.3 | - | - | - | par | $H_2O$, meth. masers | Fig. 4 in Wu et al (2014) - BeSS |
| - | - | - | - | -9.9 | - | - | par | $H_2O$ masers | Fig.15 in Choi et al (2014) - BeSS |
| - | - | - | -6.2 | - | - | - | par | $H_2O$ masers | Fig. 6 in Chibueze et al(2014)-VERA |
| - | - | -12.5 | -9.4 | -15.2 | -13.3 | - | par | $H_2O$, meth. masers | Tab. 1 in Bobylev & Bajkova (2013) |
| - | - | - | - | -9.5 | - | - | par | $H_2O$ masers | Fig. 10 in Zhang et al (2013) -BeSS |
| - | - | - | - | -17.8 | -11.6 | - | par | $H_2O$ masers | Fig. 3 in Sakai et al (2012) - VERA |
| - | - | -7.0 | -8.0 | -13.0 | -12.0 | - | par | $H_2O$, meth. masers | Fig.4 in Reid (2012) |
| - | - | - | - | - | -12.1 | - | pal | $H_2O$ masers | Fig. 5 in Sanna et al (2012) |
| - | - | -14.2 | - | - | - | - | kin | CO clouds | Fig.4 in Dame & Thaddeus (2011) |
| - | - | - | - | -12.0 | -12.6 | -5.6 | kin | HII regions | Fig. 7 in Foster & Cooper (2010) |
| - | -13.5 | -15.6 | -13.6 | -13.5 | - | - | kin | FIR [CII] & [NII] | Tab. 3 in Steiman-Cameron et al (2010) |
| - | - | - | -11.2 | - | - | - | par | $H_2O$ masers | Fig. 6a in Sato et al (2010) - VERA |
| - | -9.2 | -12.5 | -11.1 | -8.4 | - | - | kin | HII and GMC | Tab. 1 in Hou et al (2009) |
| - | - | - | - | -16.5 | -2.3 | - | par | $H_2O$, meth. masers | Fig. 2 in Reid et al (2009) |
| - | -12 | -13 | -12 | -12 | -12 | - | - | Median value | |
| - | -11.9 | -13.3 | -11.4 | -12.2 | -11.3 | - | - | Mean value | |
| - | ±1.2 | ±0.9 | ±0.9 | ±0.9 | ±1.2 | - | - | Standard dev. of the mean | |
| - | - | -14.9±5.4 | -9.7±4.1 | -13.0±3.2 | -11.4±3.9 | - | - | Mean and r.m.s. for Parallax method | |
| - | -13.7±2.4 | -13.2±1.0 | -14.0±1.0 | - | - | - | - | Mean and r.m.s. for twin-tangent arm method | |

Notes:

(a): Kinematic method (kin), Parallax method (par), Luminosity-period Cepheid method (lum), Twin-tangents method (tan), or photometric method (phot).

(b): GMC = Giant Molecular Clouds; HII = HII regions; meth. = methanol

(c): Those part of surveys are identified thus VERA (VLBI Exploration of Radio Astrometry) and BeSS (Bar and Spiral Structure Legacy survey).

**Table 2 – Recent observational studies of the spiral arms in the Milky Way (2015-early 2017).**

| Pitch Angle (deg.) | No.of arms m | Arm sha-pe [(a)] | Inter-arm [(b)] (kpc) | Data used | weight (s=strong; p=partial) | Figures and references |
|---|---|---|---|---|---|---|
| -13 | 4 | log | 3.0 [(c)] | WR stars | p | Fig.16 in Kanarek et al (2015) |
| -6 | 2 | log | 2.8 [(c)] | open star clusters | p | Fig. 14 in Scholz et al (2015) |
| -12 | 4 | log | 2.8 [(c)] | open star clusters | p | Fig. 14d and Fig.13 in Bisht et al (2015) |
| -13 | 4 | log | 2.8 | 120 spectrosc. binaries | s | Fig. 1 in Bobylev & Bajkova (2015) |
| -13 | 4 | log | 2.8 | 168 OB stars | s | Fig. 6 in Bobylev & Bajkova (2015) |
| -12.8 | 4 | log | 2.7 | runaway stars | p | Fig.7 in Vickers et al (2015) |
| -5.0 [(d)] | 1 [(d)] | log | 3.4 [(c,d)] | open star clusters | p | Fig. 5a in Griv et al (2015a) |
| -3.8 [(d)] | 1 [(d)] | log | 3.4 [(c,d)] | masers | p | Fig.5b in Griv et al (2015a) |
| -3.1 [(d)] | 1 [(d)] | log | 2.9 [(c,d)] | HII regions | p | Fig.7b in Griv et al (2015b) |
| -12 | 4 | log | 3.5 [(c)] | kinematic clouds | p | Fig.6 in Nguyen et al (2015) |
| -12.8 | 4 | log | 3.3 | cold clumps | p | Fig.3 in Marshall et al (2015) |
| -13 | 4 | log | 3.6 [(c)] | cosmic ray electrons | p | Fig.1 in Kissman et al (2015) |
| -13 | 4 | log | 3.7 [(c)] | dusty filaments | p | Fig.13 in Li, Urquhart et al (2016) |
| 0 [(e)] | 2 | ring | 2.0 [(c)] | open clusters; OB ass. | p | Fig. 2b and 4a in Melnik et al (2016a) |
| 0 [(e)] | 2 | ring | 2.0 [(c)] | open clusters | p | Fig. 1 in Melnik et al (2016b) |
| -10.5 | 4 | log | 2.5 | GMC, CO J=1-0 gas | p | Fig.7 in McGaugh (2016) |
| -13 | 4 | log | 2.0 | HI and CO | s | Fig.7 in Nakanishi & Sofue (2016) |
| -13 | 4 | log | 3.0 [(c)] | $^{12}CO$ ; $^{13}CO$ ; $C^{12}O$ | s | Fig.4 in Du et al (2016) |
| -13 | 4 | log | 3.4 [(c, f)] | CO | p | Fig.15 in Rice et al (2016) |
| -13 | 4 | log | 3.2 [(c, f)] | $^{13}CO$; mol. filaments | p | Fig.2 in Abreu-Vicente et al (2016) |
| - | 4 | log | 3.2 [(c, g)] | star forming regions | p | Fig.2 in Li et al (2016) |
| -14 | 4 | log | 3.2 [(c)] | CO 1-0 and HI 21cm | p | Fig.1 and Fig.11 in Koda et al (2016) |
| -13 | 4 | log | 2.0 | Hmol /(Hatom + Hmol) | p | Fig.11b in Sofue & Nakanishi (2016) |
| -11 | 4 | log | 3.3 [(c)] | water & meth. masers | p | Fig.1 in Reid et al (2016) |
| -12.8 | 4 | log | 3.1 | ubvi of OB star clusters | p | Fig.13 in Molina-Lera et al (2016) |
| -12 | 4 | log | 3.3 [(c)] | cold dust clumps | p | Fig.4 in Zhang et al (2016) |
| -13.1 | 4 | log | 3.1 | CO longit. of tangents | s | Fig. 2 and Sect. 3.2 in Vallée (2016a) |
| -15 | 4 | log | 3.0 [(c)] | HII regions and dust | p | Fig.11 in Bland-Hawthorn & Herhard (2016) |
| -13 | 4 | log | 3.3 [(c)] | cold dust filaments | p | Fig. 4 in Wang et al (2016) |
| -12 | 4 | log | 2.9 | nearby young stars | p | Fig. 3 in Giorgi et al (2016) |
| -13 | 4 | log | 3.0 | 79 open clusters | p | Fig.7 in Reddy et al (2016) |
| -12 | 4 | log | 2.9 [(c)] | 36 dark molecular gas | p | Fig. 3 in Tang et al (2016) |
| -13 | 4 | log | 3.1 | galactic cold clumps | s | Fig.11 in Ade et al (2016) |
| -11 | 4 | log | 3.1 [c] | local high-mass star | s | Fig.2 in Xu et al (2016) |
| -13 | 4 | log | 2.9 [c] | methanol masers | p | Fig. 5 in Hu et al (2016) |
| | | | | | | |
| -13 | 4 | log | 2.8 [c] | Open star clusters | p | Fig.5 in Bobylev et al (2016) |
| -13 | 4 | log | 3.0 | mini-starbursts clouds | p | Fig.2 in Nguyen-Luong et al (2016) |
| -1 [d] | 1 [d] | log | - | 223 Cepheids near Sun | p | Table 1 in Griv et al (2017) |
| -13 | 4 | log | 2.8 [c] | 8107 CO molecular clouds | p | Fig.10 in Miville-Deschênes et al (2017) |
| -10.5 | 4 | log | 2.5 [c] | 131 masers | s | Fig.7 in Rastorguev et al (2017) |
| | | | | | | |
| -10.5 | 4 | log | 2.5 [c] | 189 pulsar DM and RM | s | Fig.9 in Yao et al (2017) |
| -12 | 4 | log | 3.2 [c] | molecular clouds | p | Fig.4 in Watson & Koda (2017) |
| -13 | 4 | log | 3.5 [c] | mid-infrared clumps | p | Fig.3 in König et al (2017) |
| -13 | 4 | log | 2.9 [c] | Far infrared clumps | p | Fig. 1 in Veneziani et al (2017) |
| -13 | 4 | log | 2.8 | open star cluster | p | Fig.15 in Monguio et al (2017) |

| | | | | |
|---|---|---|---|---|
| -13 | 4 | log | 3.0 | Median value (rows 1-15; excluding the m=1 model and m=2 model) |
| -12.7 | - | log | 3.1 | Unweighted mean (rows 1-15; excluding the m=1 and the m=2 model) |
| ±0.3 | - | - | ±0.2 | Standard deviation of the mean (rows 1-15) |
| | | | | |
| -13 | 4 | log | 3.1 | Median value (rows 16-30; all data) |
| -12.6 | - | log | 2.9 | Unweighted mean (arrows 16-30; all data) |
| ±0.4 | - | - | ±0.2 | Standard deviation of the mean (rows 16-30) |
| | | | | |
| -13 | 4 | log | 2.9 | Median value (rows 31-45;excluding the m=1 models) |
| -12.5 | - | log | 3.0 | Unweighted mean (rows 31-45; excluding the m=1 models) |
| ±0.3 | - | - | ±0.2 | Standard deviation of the mean (rows 31-45) |

Notes:

(a): The arm shape can be logarithmic (log), ring, polynomial (polyn), irregular (irreg), or complex (comp).

(b): The separation between the Perseus arm and the Sagittarius arm, through the Sun's location.

(c): Corrected for 8.0 kpc as the Sun - Galactic Center distance (not the 8.5 value or else as used).

(d): A collection of waves, each with a different no. of arms, is proposed; only the 1-arm model is pictured.

(e): A collection of an ascending segment of pitch angle +6$^{o}$ (Perseus) and a descending segment of pitch angle -6$^{o}$ (Sagittarius) is proposed.

(f): Adapted (re-scaled) from an earlier published model.

(g): A pitch angle of 12.5$^{o}$ was imposed, in order to deliver a good fit to the observed data.

**Table 3. Some published velocimetric models for the spiral arms of the Milky Way**

| Model | $R_{sun}$ used (kpc) | Mean $V_{lsr}$ used (km/s) | Arm pitch angle ($^{O}$) | Arm starts from GC (kpc) | Arm shape | Gal. Qua-drants used |
|---|---|---|---|---|---|---|
| Shane (1972) | 10.0 | 250 | -7.8 | 3.5 | log | I |
| Vallée (2008) | 7.6 | 220 | -12.8 | 3.1 | log | I,II,III,IV |
| Dame & Thaddeus (2011) | 8.5 | 220 | -14.2 | 4.0 | log | I |
| Sanna et al (2014) | 8.4 | 243 | -12 | 4.0 | log | I, IV |
| Reid et al (2014) | 8.3 | 240 | -12.5 | 3.8 | log | I,II |
| Reid et al (2016) | 8.3 | 240 | -11 | 3.0 | log | I,IV |
| Green et al (2017) | 8.4 | 246 | -12 | 3.1 | log | I, IV |
| Vallée (2017b,c) | 8.0 | 230 | -13.1 | 2.2[( a)] | log | I,IV[( b)] |

Note a: All velocimetric models start their arm between 3 and 4 kpc from the Galactic Center, except our own model starting at 2.2 kpc. Thus, our model allows the Norma arm in Quadrant I to reach higher velocities near the GC (Vallée 2017b).

Note b: We have employed this velocimetric model here, in order to cover the Galactic Quadrants II and III (fig. 5b).

**Figure captions**

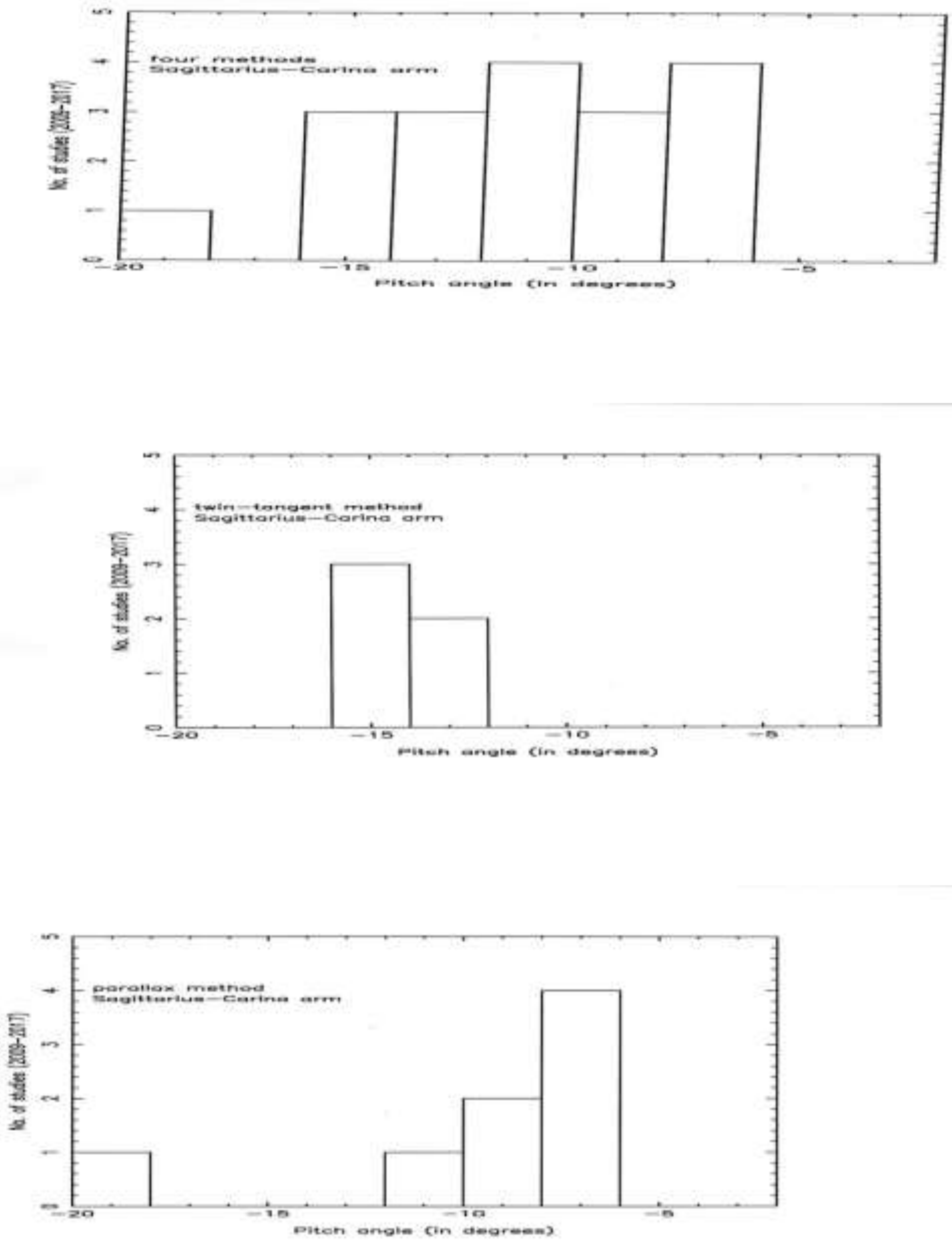

Figure 1. Histogram of the observed individual arm pitch angle values, for the Sagittarius-Carina arm, from 2009 to early 2017.

a) all data.

b) data from twin-tangent arm method.

c) data from the parallax method.

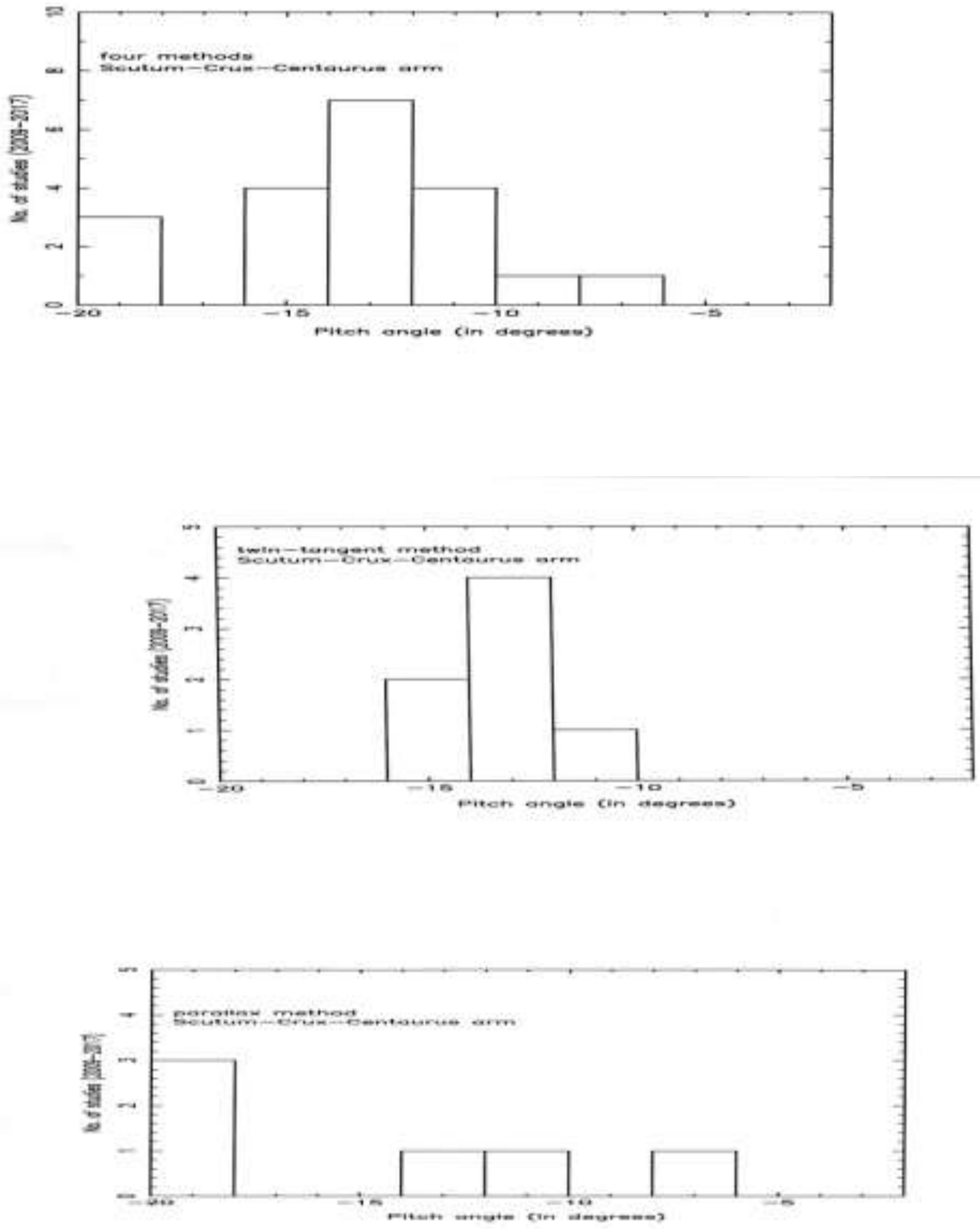

Figure 2. Histogram of the observed individual arm pitch angle values, for the Scutum-Crux-Centaurus arm, from 2009 to early 2017.

a) all data.

b) data from twin-tangent arm method.

c) data from the parallax method.

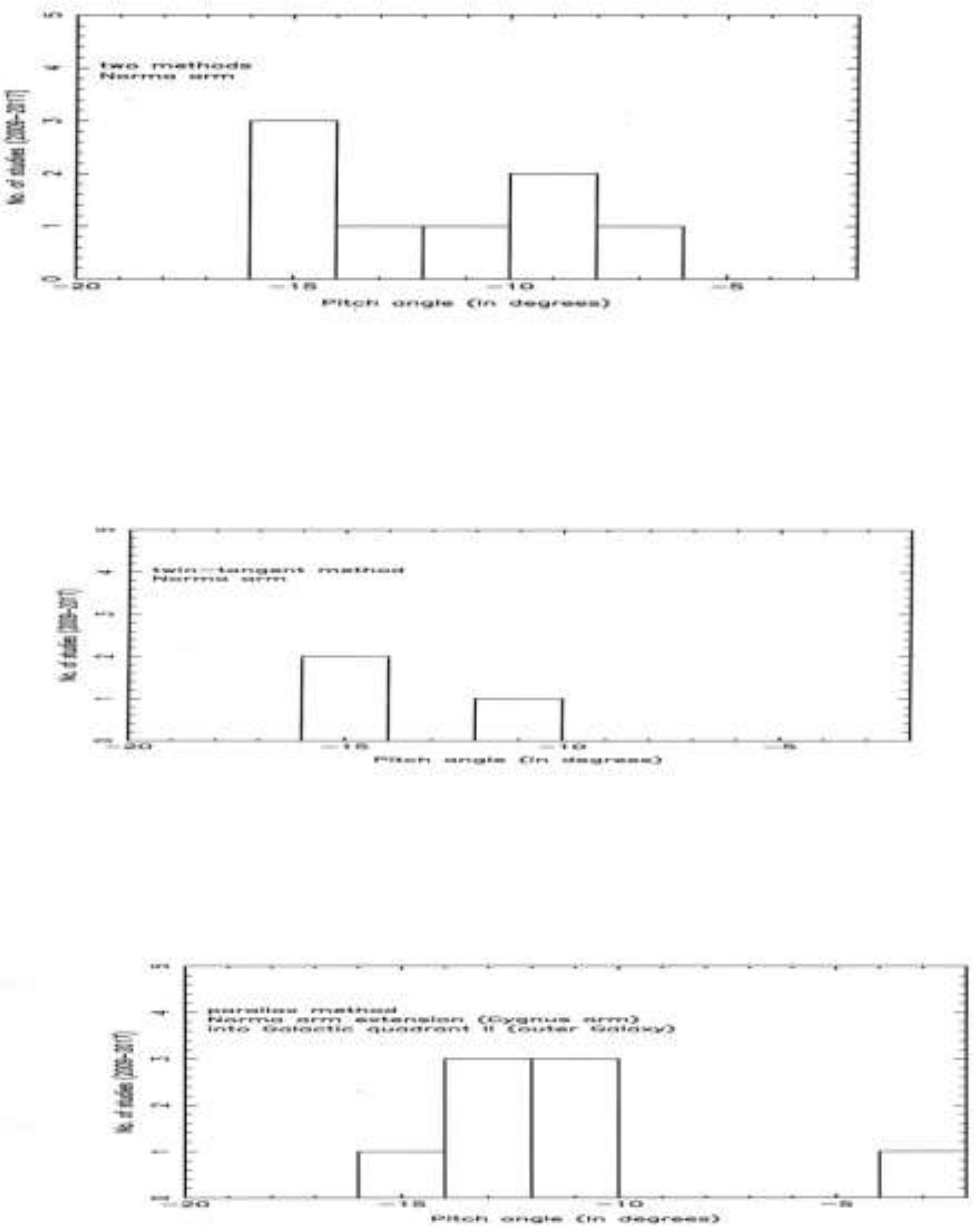

Figure 3. Histogram of the observed individual arm pitch angle values, for the Norma arm, from 2009 to early 2017.

a) all data.

b) data from twin-tangent arm method.

c) data from the parallax method for the Norma arm extension in Galactic Quadrant II (so-called the Cygnus arm).

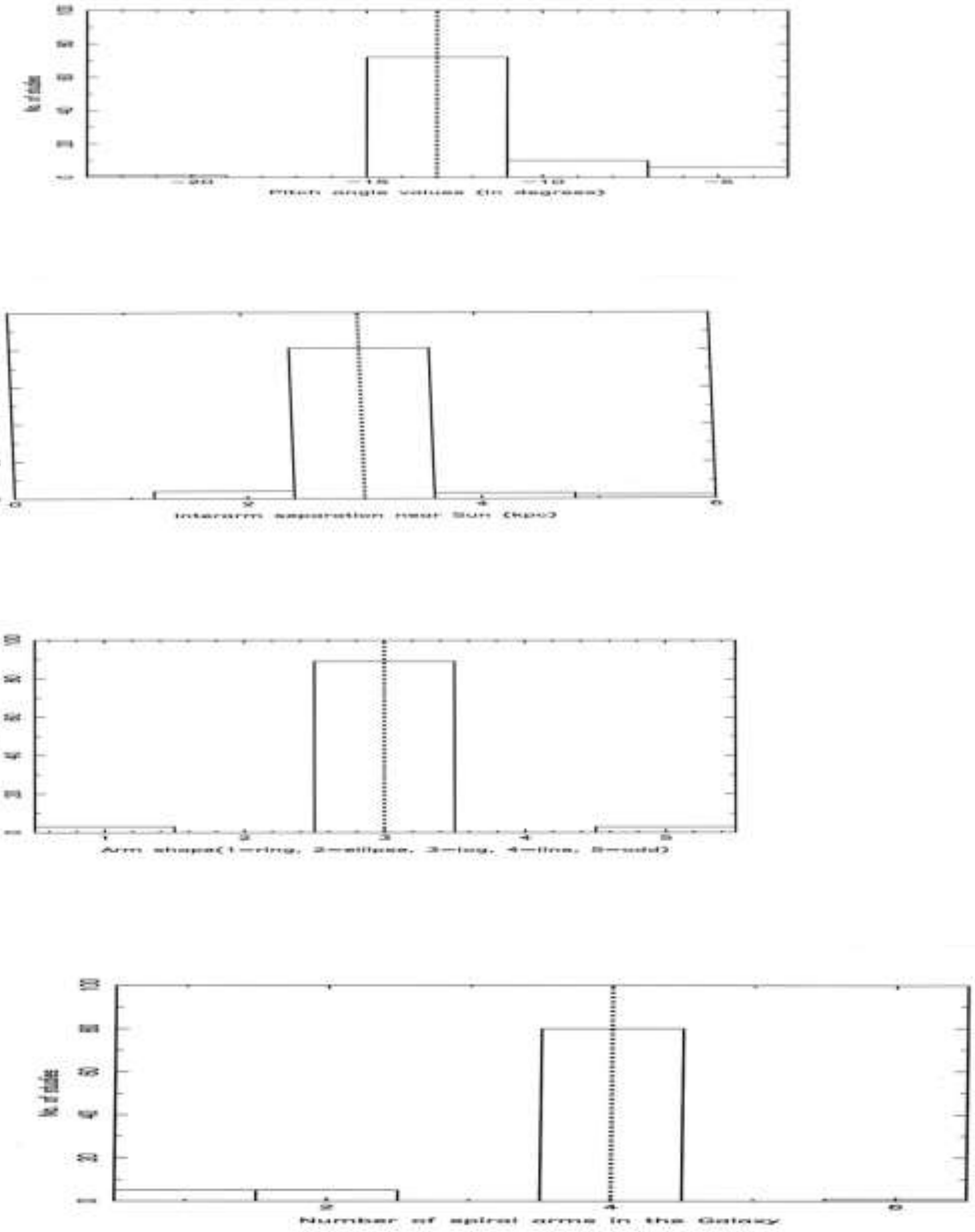

Figure 4. Histograms for the published positional studies since 2014.

a) Histogram of global spiral arm pitch angle.

b) Histogram of the interarm separation, across the Sun's position.

c) Histogram of global spiral arm shape.

d) Histogram of the number of spiral arms.

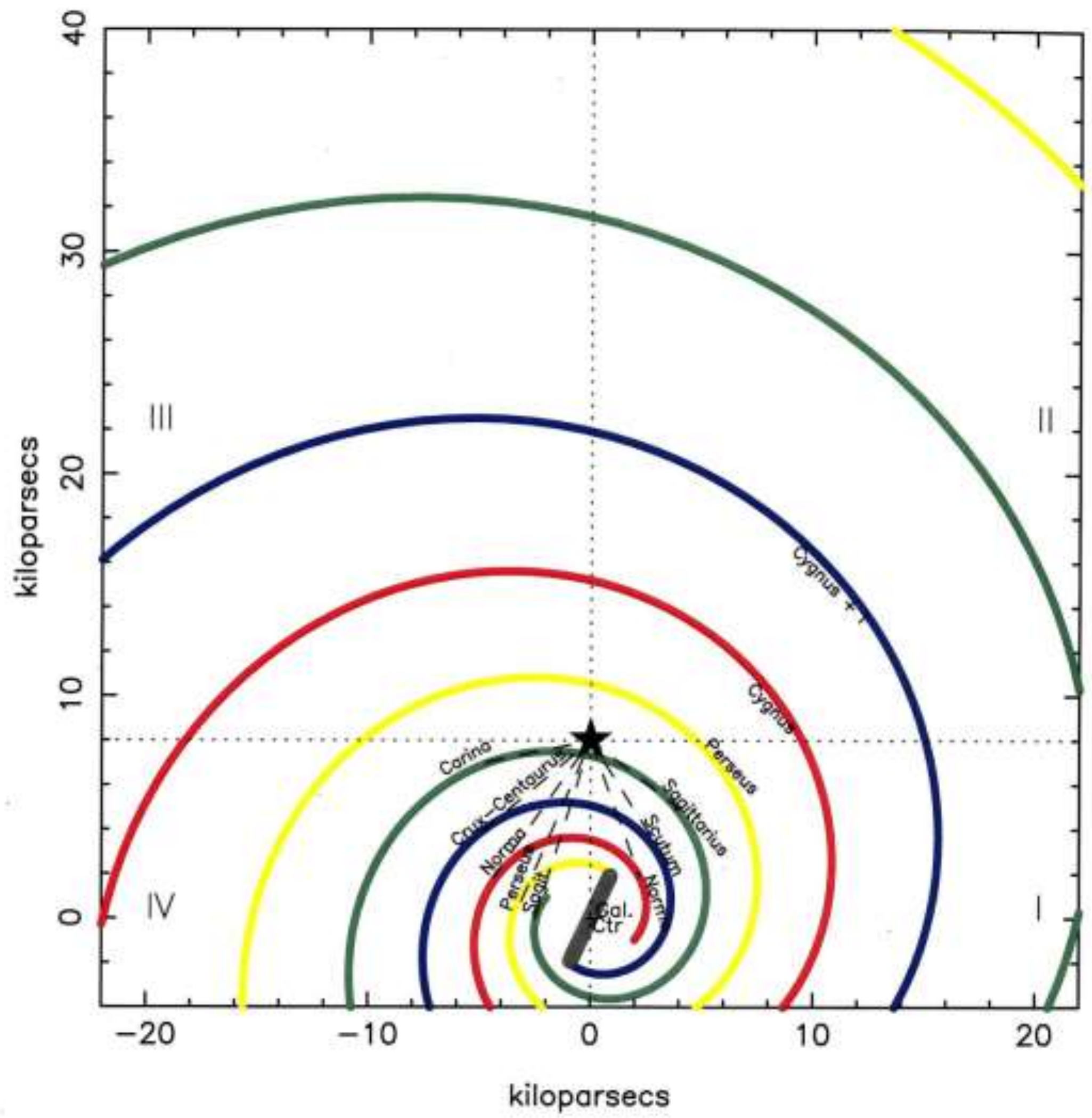

Figure 5a. Rough cartographic model of the face-on map of the Milky Way disk, showing and naming the individual arms, centered on the Outer Galaxy, North of the Sun ($90^{o}$ < galactic longitude < $270^{o}$). For an expanded version of the model South of the Sun, see Fig.3 in Vallée (2017c).

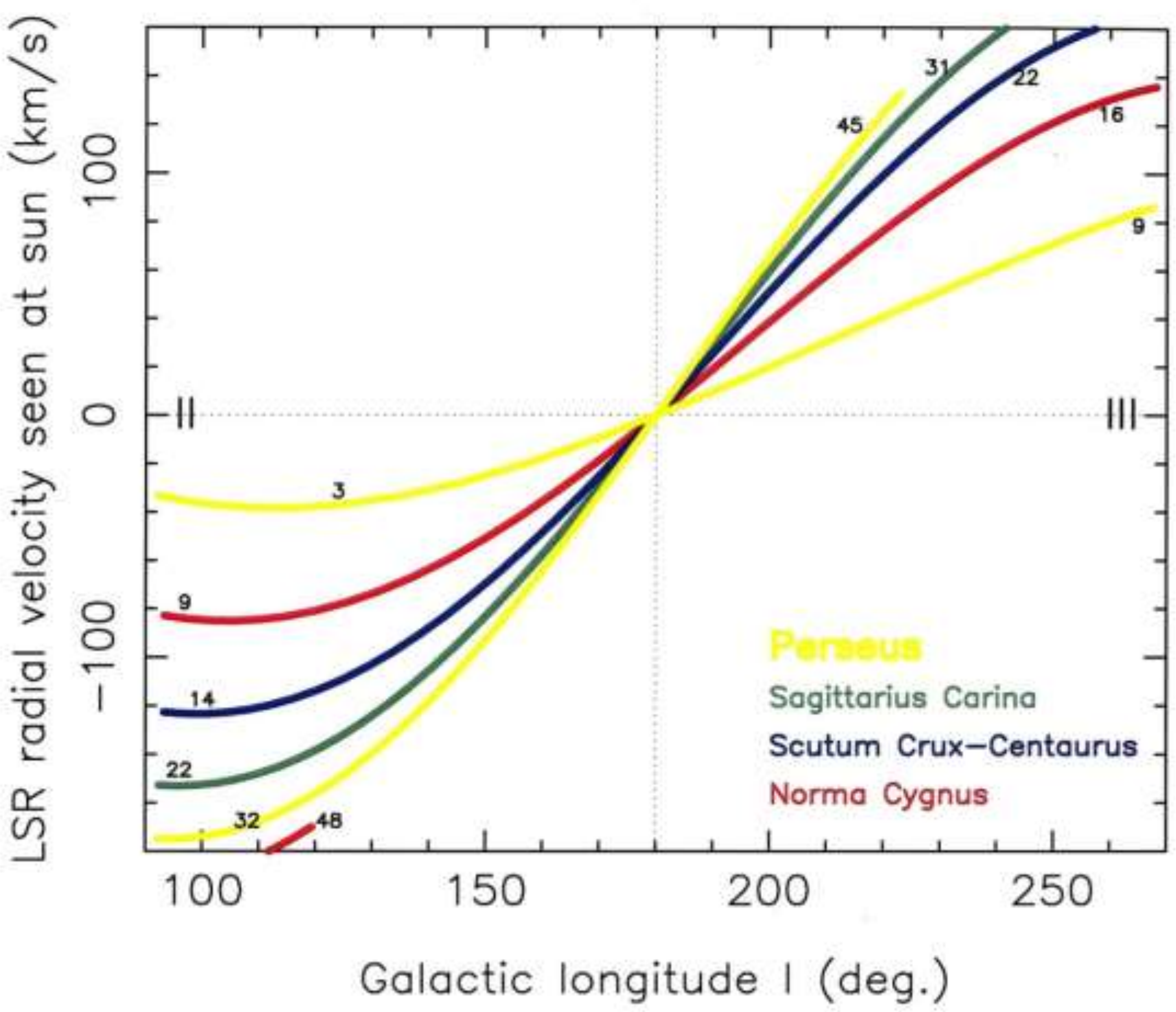

Figure 5b. Rough velocimetric map of the Outer Milky Way galaxy, centered on Galactic quadrants III and II ($90^o$ < longitude < $270^o$). The number on each arm indicates the rough distance to the Sun (in kpc). For the model in the other Quadrants I and IV, see Fig. 4 in Vallée (2017c).